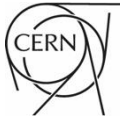

# FIRST PROOF-OF-CONCEPT PROTOTYPE OF AN ADDITIVE-MANUFACTURED RADIO FREQUENCY QUADRUPOLE

*T. Torims[1,2], G. Pikurs[1,2], S. Gruber[3], M. Vretenar[2], A. Ratkus[1,2], M. Vedani[4], E. Lopez[3], F. Bruckner[3,5]*

[1] Riga Technical University, Riga, Latvia
[2] CERN, Geneva, Switzerland
[3] Fraunhofer Institute for Material and Beam Technology, Dresden, Germany
[4] Politecnico di Milano, Milan, Italy
[5] Luleå University of Technology, Sweden

## Abstract

Continuous developments in Additive Manufacturing (AM) technologies are opening opportunities in novel machining, and improving design alternatives for modern particle accelerator components. One of the most critical, complex, and delicate accelerator elements to manufacture and assemble is the Radio Frequency Quadrupole (RFQ) linear accelerator, used as an injector for all large modern proton and ion accelerator systems. For this reason, the RFQ has been selected by a wide European collaboration participating in the AM developments of the I.FAST (Innovation Fostering in Accelerator Science and Technology) Horizon 2020 project. RFQ is as an excellent candidate to show how sophisticated pure-copper accelerator components can be manufactured by AM and how their functionalities can be boosted by this evolving technology. To show the feasibility of the AM process, a prototype RFQ section has been designed, corresponding to one-quarter of a 750 MHz 4-vane RFQ, which was optimised for production with state-of-art Laser Powder Bed Fusion (L-PBF) technology, and then manufactured in pure copper. To the best knowledge of the authors, this is the first RFQ section manufactured in the world by AM. Subsequently, geometrical precision and surface roughness of the prototype were measured. The results obtained are encouraging and confirm the feasibility of AM manufactured high-tech accelerator components. It has been also confirmed that the RFQ geometry, in particular the critical electrode modulation and the complex cooling channels, can be successfully realised thanks to the opportunities provided by the AM technology. Further prototypes will aim to improve surface roughness and to test vacuum properties. In parallel, laboratory measurements will start to test and improve the voltage holding properties of AM manufactured electrode samples.

# Contents

# 1        Additive Manufacturing for the RFQ

Currently the Radio Frequency Quadrupole (RFQ) is a core element for hundreds of industrial and research linear accelerators operating in the world. The RFQ is a compact and sophisticated accelerator which simultaneously focuses, bunches, and accelerates a continuous beam of positively charged particles like protons or heavier ions as they come out of the ion source [1]. Today RFQ's are conventionally manufactured from highly conductive (e.g. oxygen-free high thermal conductivity - OFHC copper) materials and alloys. After decades of test-and-trial, current manufacturing technology for RFQ's of the 4-vane type consists of multi-axis high-precision milling of pre-fabricated large-scale-forged single-piece components (see Fig. 1). The full RFQ consists of four modules with complex and high-tolerance manufactured surfaces that are subsequently joined together in the final "4-vane" configuration by furnace brazing [2]. This last technological process often releases residual stresses that may result in excessive geometrical distortion and disqualifies the end-product from further use. A complex procedure requiring several stress release thermal treatments during machining is required to ensure that the tight tolerances are respected after brazing. This results in a costly, time consuming and inefficient process with an extremely high rate of waste material rate. Furthermore, RFQ's normally go through further machining steps to realise long and complex internal cooling channels and ports for slug tuners, monitoring loops, vacuum pumps, and RF couplers.

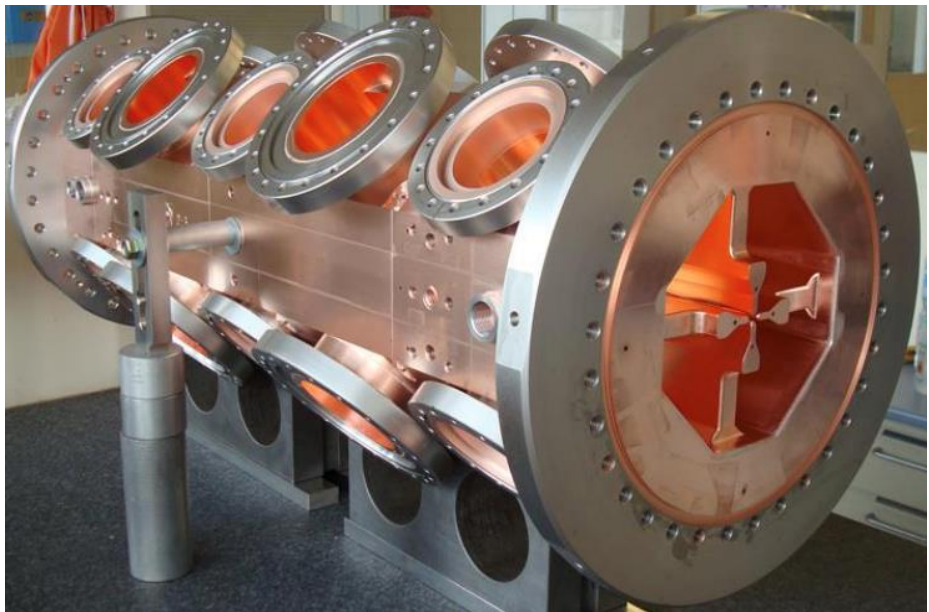

**Fig. 1:** The CERN's LINAC 4 RFQ module design (CERN) [3]

Virtues of the Additive Manufacturing (AM) and the latest developments of the technology are particularly well placed to improve the manufacturing aspects of the RFQ, promising to significantly reduce machining time and costs as well as to realise an improved design. Eventually, complete segments including all four "vanes" of the RFQ system could be built in one piece, thus avoiding brazing, and allowing for the optimal manufacturing of complex elements as internal cooling channels and external ports. Advances in AM equipment, design ability (including simulation tools) and the manufacturing methodology itself are opening entirely new avenues for the RFQ design optimisation and full-scale production, even using pure-copper, which is considered as a challenging material for laser-based AM processes. Naturally, this is well suited for the needs of the particle accelerator community and RFQ manufacturing in particular.

This paper outlines, to the best of authors' knowledge, the very first proof-of-concept confirming that AM manufactured RFQ is feasible and achievable. At the same time, it acknowledges key technological features, which have to be addressed in order to widen and proliferate AM technology in the accelerator community.

## 1.1 RFQ specific requirements

RFQ is a component of particle accelerators featuring strict technical requirements for its successful field service. At the first glance, it appears that stringent requirements (see Table 1) are almost unreachable by current state-of-art of the AM systems. However, the continuous development of AM systems and related post processing technologies are steadily approaching the required level of RFQ precision and surface quality as well as manufacturing predictability. The experimental testing activity of this proof-of-concept was performed on the commercially available state-of-art laser-based AM technology, suitable for the pure copper manufacturing. Table 1 summarises the main parameters of the design and manufacturing of pure copper RFQ.

The manufacturing experiment was carefully designed and planned, keeping in mind the requirements of Table 1. To ensure the functionality of the RFQ, geometrical accuracy and shape of the manufactured surfaces is of utmost importance, as indicated by the values of 20 µm on vane tip and 100 µm for all other surfaces. The most relevant target value here is the RFQ vane tip and its modulation profile, which is the core element for beam transport – therefore particular attention and measurements will be devoted to the vane tip. Clearly, if one cannot provide enough precision on the modulation geometry, beam transport and acceleration can-not be ensured.

**Table 1:** Requirements for the prototype RFQ

| Requirement | Target values |
|---|---|
| Geometrical accuracy | 20 µm on vane tip, 100 µm elsewhere |
| Surface roughness | $R_a$=0.4 µm for all inner surfaces |
| Porosity, degassing | Vacuum $10^{-7}$ mbar |
| Electrical conductivity | 90% as per International Annealed Copper Standard |
| Peak electric field on surface | ~ 40 MV/m |

Furthermore, surface arithmetical mean roughness value $R_a$ has to be kept at level of about 0.4 µm. Surface roughness has to be smaller than the penetration of high-frequency currents in the metal ("skin depth") to avoid considerable reductions in the Q-value of the RFQ resonator and a proportional increase in its power consumption and in the cost of the Radio-Frequency system. Moreover, large values of $R_a$ might increase the sparking probability of surfaces subject to high electric fields. Although surface roughness is critical for the functionality of RFQ, such values are rather difficult to maintain with conventional AM technology and might require post processing of the surfaces transporting the radio-frequency current.

The vacuum value of $10^{-7}$ mbar is set as a minimum required value for the RFQ – circular accelerators often require lower pressures.

The electrical conductivity is of utmost importance and has a decisive impact on RFQ efficiency. The highest electrical conductivity can be reached only with high chemical purity and density of the base material – e.g. copper. In the case of AM, the chemical purity of the final product depends not only on chemical cleanness of powder, but also on the manufacturing chamber protection against oxidation. It is important to note that the oxygen-free pure copper powder grains tend to oxidize already at standard room environment and temperatures. Lower electrical conductivity of the RFQ in turn will proportionally increase the required operational power of the accelerator, in a similar way to the roughness, and will generate extra heat on the vane surfaces. Therefore, target value for the electrical conductivity for this proof-of-concept is set to 90% of ideal copper according to the International Annealed Copper Standard (IACS).

Finally, the voltage holding properties are crucial for the successful operation of the RFQ. Naturally, these properties are directly affected by any mechanical and chemical inclusions as well as homogeneity of the RFQ material itself. Considering some existing RFQ design, a target value can be empirically defined at about 40 MV/m peak surface field.

However, it was clear that not all RFQ specific requirements could be achieved at this initial proof-of-concept stage (e.g. roughness, degassing and voltage holding). In the proposed prototype, design emphasis is given to the verification of AM capabilities for the RFQ geometrical accuracy (manufacturing tolerances), surface quality (roughness) as well as to the demonstration of improved mechanical design advantages.

## 1.2    AM technology and challenges

AM processes for metals can be divided into nozzle based processes and powder bed based processes. Nozzle based processes feed the raw material, powder or wire, through a nozzle to the work zone into the focus of an energy source which can be a laser, electron beam or an electric arc. Powder bed processes either use a laser (Laser Powder Bed Fusion – L-PBF) or electron beam (Electron Beam Melting – EBM) as an energy source or a binder (Binder Jetting - BJ) to fuse the powder together. L-PBF is the most promising AM process for pure copper RFQ, thanks to the fact that a) high relative density and high electrical conductivity can be achieved, and b) build-up of complex-shaped parts is possible with a minimum wall thickness of 400 microns and a layer thickness of 30 microns. These challenging material properties can be attained by deploying a short wavelength laser because the absorptivity level of the pure copper is very low within the commonly used infrared L-PBF systems and significantly increases in the green wavelength. Thus, the energy coupling into the pure copper powder bed increases, and defect-free processing is possible by using the green laser source [4,5]. At the same time, the L-PBF technology is well-placed for the required mechanical complexity and offers significant design and optimisation freedom to meet the requirements for the RFQ (i.e. integrated cooling channels) that cannot be achieved by the mentioned nozzle technologies [4–8].

Nonetheless, there are still some remaining issues of the L-PBF process to overcome, such as design restraints regarding the minimum wall thickness or maximum overhang angle without support and tolerance specifications [9]. The minimum overhang angle is 45° and the minimum wall thickness for this material and machine used (see section 3.1) in this proof-of-concept is 0.6 mm. The surface roughness, tolerances and geometry of the RFQ are rather demanding and cannot be ensured *per-se* by the L-PBF standard process due to the staircase effect, adhesion of powder particles and material distortion during the cooling of the part. Therefore, at the outset of this proof-of-concept it is evident that the whole process chain of RFQ manufacturing with L-PBF will require future improvements and the fine-tuning of the technological process itself and eventually may require subsequent post-processing stages.

The removal of powder can also be critical when using internal cavities. In the case of the proof-of-concept, to ensure that all residual manufacturing powder is eliminated, the prototype was cleaned with pressurized air and in an ultrasonic bath.

# 2        Optimisation of prototype RFQ

## 2.1    Design improvements

The design of the proof-of-concept RFQ is intended to reproduce one quarter of CERN's high frequency (HF) RFQs recently built for applications in the medical and detection fields (see Fig. 2) [10], however, only vane tip geometry and main geometrical-shape proportions were kept unchanged. Most of the external and internal shapes have been optimized exclusively for AM, considering its advantages and opportunities as well as its requirements and restrictions. At the design development stage, a multidisciplinary team of accelerator physicists, manufacturing technologists and AM experts was established to find the optimal and balanced technological solution, taking into account potential manufacturing time and cost, geometrical precision, surface parameters, structure rigidity and thermal stability.

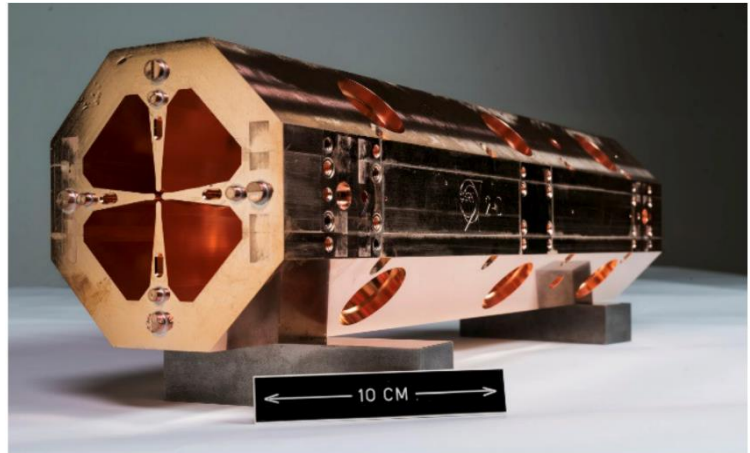

**Fig. 2:**  The CERN's 750 MHz RFQ module (CERN) [10]

As a result of the extensive teamwork, a model of the RFQ quarter with a length of 95 mm was designed, based on AM manufacturing capabilities, state-of-art equipment (see section 3) and bearing in mind the high material costs. The RFQ has a quadrupolar symmetry and therefore all complexity and characteristic elements can be encompassed within the 90° RFQ sector (see Fig. 3) which is an indicative sample of the whole optimised RFQ structure. The quarter-sector prototype includes the vane tip, inner surfaces, improved cooling channels and re-designed inner structure.

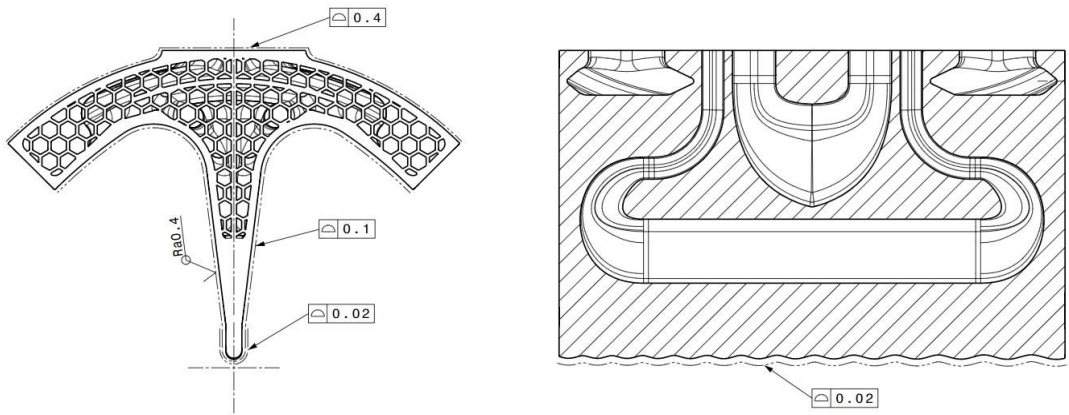

**Fig. 3:** RFQ quarter-sector - design for AM with improved cooling channels system (G.Pikurs, CERN)

The lightened structure of RFQ was re-designed by using a honeycomb pattern (see Fig. 3 and 4). Thus replacing the most massive sections and introducing shaped cooling channels as in Fig. 3 results in a significant material volume reduction of ~37% in comparison with conventional RFQ designs. The honeycomb structure and cooling channel re-design reduced the weight by 21% and 16% respectively. A honeycomb structure with a wall thickness of 0.6 mm was chosen; this value is slightly above the minimum wall thickness for a reliable manufacturing process. Naturally, all cooling channel shapes and AM build inclination angles were adapted for AM requirements. The thermal analysis, which is described in the next section also was taken into account for design development.

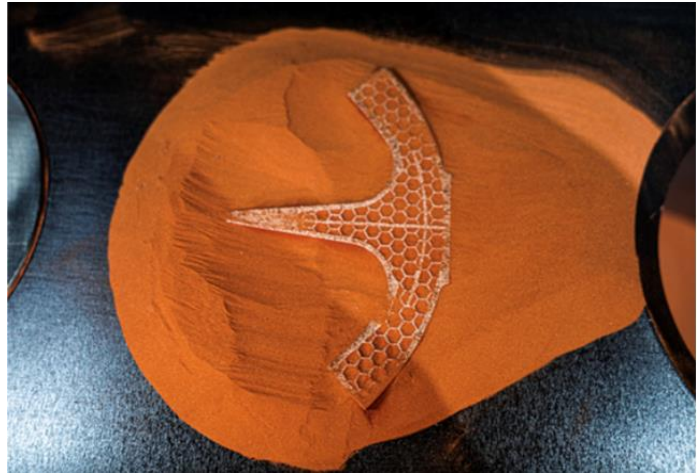

**Fig. 4:** RFQ proof-of-concept prototype - build job, build plate at bottom position (S.Gruber, Fraunhofer IWS)

## 2.2    Thermal analysis

The basic concept of the thermal management for AM produced RFQ was tested on Ansys 19.1 Steady-State Thermal analysis workbench. Input data for ANSYS simulation are based on general approximations and assumptions from the recently built at CERN 750MHz PIXE RFQ [11]. Crucial input data for the analysis were: a) 22°C cooling channel temperature, b) heat flux on vane tip $2x10^{-3}$ W/mm$^2$, c) flux on the vane and internal walls $8x10^{-3}$ W/mm$^2$, and additional negligible values for heat loss through convection from outer surfaces. Thermal analysis results are provided in Fig. 5. From the Steady-State Thermal analysis it is evident that the difference of 0.8 °C is not posing any risk for the RFQ functionality. The proposed design concept, especially honeycomb internal structure and improved cooling channels, could be highly beneficial for the AM manufactured RFQs as well to other complex-shape and structure accelerator components.

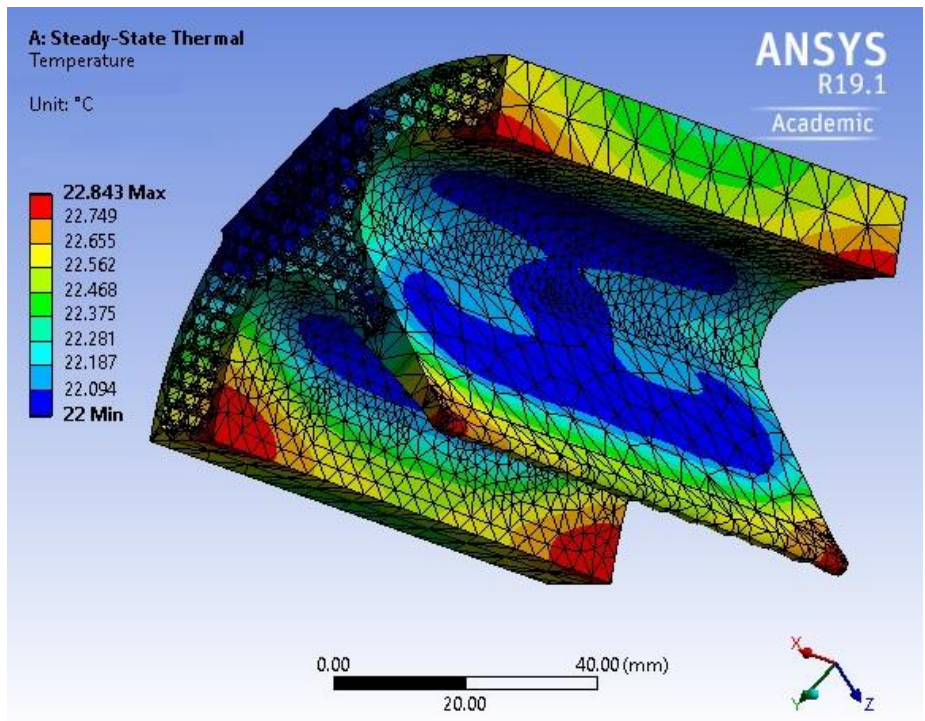

**Fig. 5:** RFQ proof-of-concept prototype - Steady-State Thermal analysis (G.Pikurs, CERN)

## 3      Manufacturing

### 3.1    AM specific needs and optimisation of the process parameters

In order to attain the best possible results, state-of-art AM technology and manufacturing equipment has been chosen for the production of the very first pure-copper RFQ prototype. A TruPrint1000 Green Edition (see Fig. 6) in combination with a green TruDisk1020 laser providing the wavelength of 515 nm and maximum laser power of 500 W was used at Fraunhofer IWS in Dresden. This machine allows for the cylindrical build volume of 100 mm in diameter and 100 mm in height. The last one was used as a limiting technological parameter for the proof-of-concept RFQ height. The TRUMPF pre-set pure-copper manufacturing technological parameters and algorithms were used throughout carefully monitored machining process.

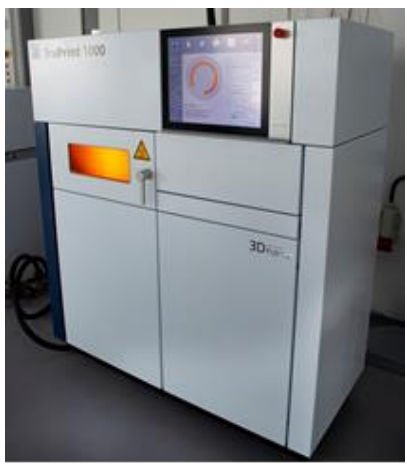

**Fig. 6:** TruPrint1000 Green Edition at Fraunhofer IWS

As a production material, m4p™ PureCu gas-atomized spherical shaped powder was used, which was confirmed with the Camsizer X2 and dynamic imaging analysis (see Table 2 and Fig. 7), measuring the sphericity at 0.923 and through scanning electron microscope imaging. The particle size distribution was confirmed between 19.5 and 34.9 µm which is common for L-PBF processes.

**Table 2:** Main characteristics of particle size distribution of Cu-ETP

| Powder | D10 in µm | D50 in µm | D90 in µm | Sphericity |
|---|---|---|---|---|
| Cu-ETP (Electrolytic Tough-Pitch: pure copper) | 19.5 | 26.2 | 34.9 | 0.923 |

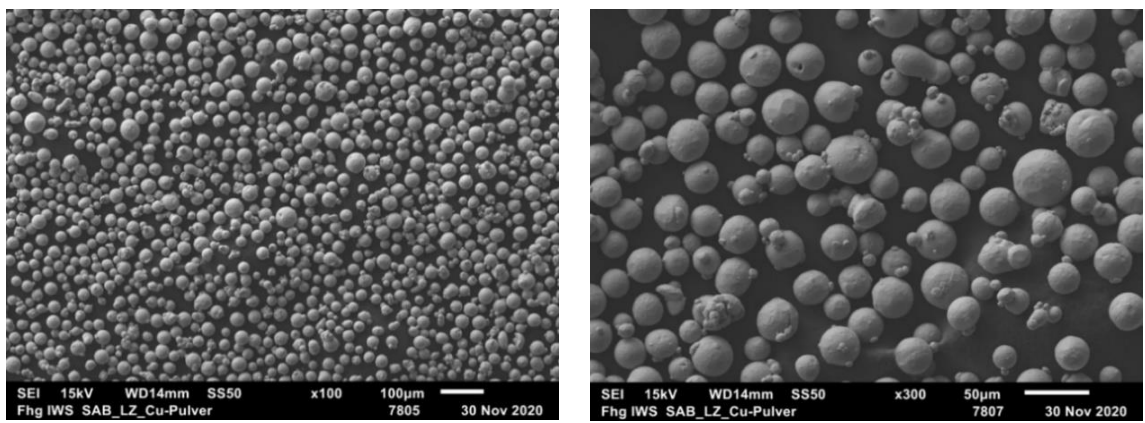

**Fig. 7:** SEM images of Cu-ETP powder, left: magnification of x100, right: magnification of x300

The used process parameters were based on previous investigations published by Gruber et al. [4], who qualified the parameters set for two different pure copper powders and by Wagenblast et al. [5], who achieved sample relative densities above 99.9 % and electrical conductivity above 100 % IACS. To prepare and tune for the full prototype manufacture, several simulations and pre-trials took place and valuable experience was obtained. Further details and analysis of the processing conditions, material microstructure and other sample properties will be provided in future publications.

The printing job took 16h 29min, the build height was 98.01 mm and thus consisted of 3267 layers (for a layer thickness of 30 μm). The manufactured prototype can be seen in Fig. 8.

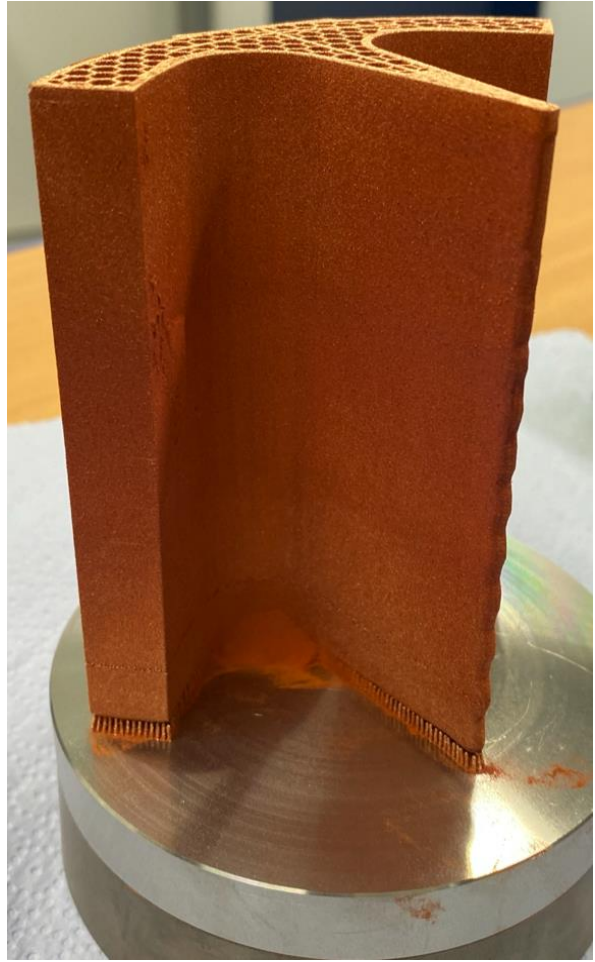

**Fig. 8:** RFQ proof-of-concept prototype – out of production and still joined to the build plate
(T. Torims, RTU/CERN)

## 4       Obtained geometrical accuracy and surface roughness

### 4.1     Geometrical accuracy

To measure the geometrical accuracy, the 3D scanner ATOS GOM Core 135 was used for a 3D optical surface scan of the as the manufactured proof-of-concept RFQ at Fraunhofer IWS (Fig. 9).

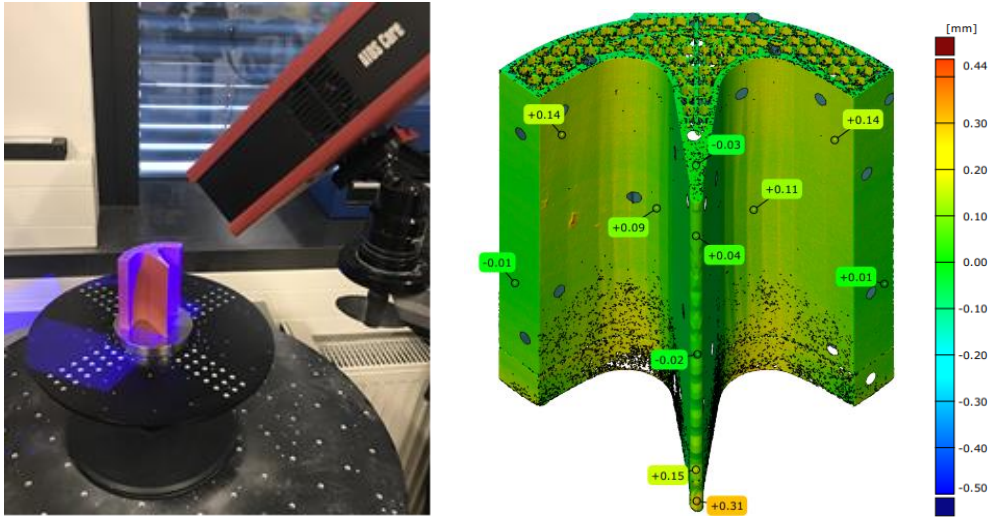

**Fig. 9:** RFQ proof-of-concept prototype - left: 3D optical scanning set-up (S.Gruber, Fraunhofer IWS), right: comparison of measured point cloud with CAD data

The largest deviation of 0.31 mm on the vane tip was found in the bottom region of the sample. The deviation probably originated due to distortion of the support structures during the build process. Therefore, in the future it is recommended to use solid supports in the region of the vane tip or throughout the bottom of the RFQ. However, the deviations along the vane tip decrease to 0.02÷0.04 mm which is a particularly promising result for the geometrical accuracy of this first prototype of AM manufactured RFQ. Furthermore, the geometrical accuracy on the outer sides of the quarter left and right of the RFQ were only ±0.01 mm.

## 4.2    Surface roughness

The surface roughness was measured at Fraunhofer IWS with the perthometer Surfcom Touch 50 which is contact profilometer from Accretech, on both sides of the vane (see Fig. 10. measurement No. 9 and 10) and RFQ internal surfaces (see Fig. 10. measurement No. 6 and 8); repeating each measurement three times.

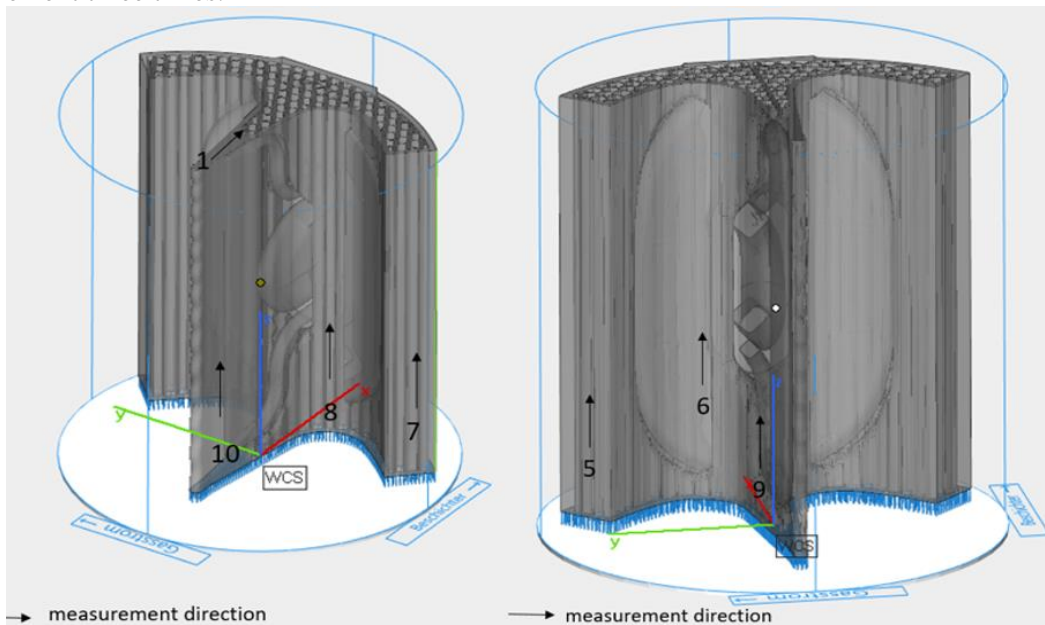

**Fig. 10:** RFQ proof-of-concept prototype - surface roughness measurement locations and direction

The surface arithmetical mean roughness $R_a$ average was calculated at 14.32 µm and the maximum height of the profile - $R_z$ at 116.7 µm. The latter was measured additionally to RFQ prototype requirements (Table 1) to obtain more comprehensive understanding of the surface quality of the prototype. The most indicative surface roughness measurement results are provided in the Table 3.

**Table 3:** RFQ proof-of-concept surface roughness values according to ISO 4288

| Location | $R_a$ (µm) | | | | $R_z$ (µm) | | | |
| --- | --- | --- | --- | --- | --- | --- | --- | --- |
| | Measurement No. | | | | Measurement No. | | | |
| | 1 | 2 | 3 | mean | 1 | 2 | 3 | mean |
| 6 | 10.4 | 12.4 | 12.8 | 11.9 | 84.2 | 89.5 | 85.6 | 86.5 |
| 8 | 15.1 | 15.0 | 15.3 | 15.1 | 148.8 | 138.7 | 143.0 | 143.5 |
| 9 | 13.8 | 14.9 | 13.5 | 14.1 | 117.2 | 123.6 | 104.7 | 115.1 |
| 10 | 13.9 | 14.9 | 14.9 | 14.6 | 117.5 | 134.2 | 103.3 | 118.3 |

Although these first roughness measurements of the proof-of-concept RFQ show that the obtained surface roughness quality is still far from the required $R_a$=0.4 µm, it is important to keep in mind that these results were obtained without any specific adaptation of the AM technological process to strike for the better surface roughness outputs. Therefore, even before considering potential post processing needs and methods for the AM made RFQ, clearly there is a range of opportunities to optimise the pure-copper L-PBF manufacturing processes itself and to attain a better surface roughness quality. This shall be part of the future experimental and research work along with the consideration of the appropriate post-processing scenarios and experiments.

## 5    Conclusions and way forward

Joint multidisciplinary effort proved that a pure-copper RFQ prototype can be successfully manufactured with AM technology, and is, to the best knowledge of authors, the first AM manufactured RFQ in the world. Indeed, the latest developments of the AM technology are providing much-needed means to improve the manufacturing aspects of the RFQ and could considerably reduce machining time and overall costs. Therefore, thanks to dedicated teamwork, the concepts on how to improve RFQ design and manufacturing features offered by the state-of-art L-PBF technology were described in this paper. The pure-copper RFQ proof-of-concept allowed the following conclusions to be derived:

1. AM technology is particularly well suited for the required mechanical complexity of RFQ and offers significant design and optimisation freedom to meet the stringent manufacturing requirements that cannot be achieved by conventional technologies. This also opens a way to major RFQ improvements and eventually a full-size production, even using pure-copper, which is a technologically demanding material.

2. Pure copper RFQ prototype, using L-PBF system equipped with a green laser, can be manufactured in reasonable time – 16h 29min with 3267 layers of 30 µm layer thickness.

3. Most of the external and internal shapes of the RFQ can be successfully optimised. The lightened RFQ structure is feasible by using a honeycomb pattern and by replacing the most massive sections.

4. Shape and structure of the RFQ cooling channels, can be improved according to the optimum thermal management and flow-dynamics needs – and not dictated by technological restrictions of the conventional manufacturing.

5. The honeycomb structure implementation and optimisation of the cooling channels are allowing for substantial weight reduction – in this case ~37% (~21% and ~16% respectively).

6. The Steady-State Thermal analysis showed that for the operating conditions of the CERN PIXE RFQ the temperature difference between different sectors remains in order of ~0.8 °C – thus not posing any risk for the RFQ functionality.

7. The surface roughness measurements indicated that the prototype surface roughness quality is still far from the required $R_a$=0.4 µm. The surface arithmetical mean roughness $R_a$ average was registered at 14.32 µm and the maximum height of the profile $R_z$ at 116.7 µm. However, these results are encouraging, since were obtained without any adaptation of the AM technological process for better surface roughness outputs.

8. The geometrical accuracy measurements revealed promising results – with the conventional AM methodology approaching the required precision of 20 µm on vane tip and fully reaching 100 µm on other surfaces. The largest deviation of 0.31 mm on the vane tip can be attributed to the technological glitch – distortion of the support structures during the build process.

## 5.1 Lessons learned

To prepare and tune for the full prototype manufacture, several simulations and pre-trials took place and valuable experience was gained throughout this proof-of-concept exercise, both for the L-PBF pure-copper technological process developments and the RFQ design improvements. There is no doubt that these encouraging proof-of-concept results could be only achieved with an open-minded and truly multidisciplinary approach to this endeavour. Importantly, this proof-of-concept project involved new partners and brought-in world-class brand-new AM expertise into the particle accelerator community. These partners learned how to work together and how to trust each-others expertise. Results are evident.

It appears that the whole process of the RFQ manufacturing with L-PBF will require additional improvements and fine-tuning. To achieve the utmost quality of the AM manufactured pure-copper RFQ, further development of the technological chain may require subsequent surface finishing steps and post processing stages. The technological and constructive optimisation shall include the modifications of the initial design (CAD part), e.g. to add an offset for the post processing as well as the use of simulation tools in order to calculate the desired pre-deformation of the RFQ surfaces, before manufacturing, to compensate the distortion. Individualized support structures in critical areas of the part can also minimize distortion.

## 5.2 Way forward

The initial proof-of-concept results presented in this paper, are clearly demonstrating the manufacturing feasibility of the complex pure copper accelerator components already with the commercially available AM solutions offered by the current state-of-art. This is already a valuable contribution to the I.FAST programme [12], and the technological approach itself could be instrumental in pushing forward accelerator technologies as requested by e.g. the updated European Strategy for Particle Physics [13].

Next steps will be to perform in-depth specialised RFQ metrological measurements following the well-established algorithms used for the HF RFQs at CERN, followed by the relevant vacuum, degassing and voltage holding tests. A sample AM manufactured piece is being designed to be tested for voltage holding at a special test bench used for CLIC RF testing at CERN.

Furthermore, based on the experience gained with the proof-of-concept RFQ, it is foreseen that a complete RFQ (all four vanes in a single piece) will be AM manufactured at Politecnico di Milano, as the next prototype. All the testing and measurement sequences will be repeated as for the proof-of-concept RFQ.

At the same time, further research and experimental work shall focus on suitable post-processing techniques and larger build volume with the L-PBF system laser configuration. There is a range of options to optimise the pure-copper manufacturing processes and potential means to attain the required surface roughness quality. This shall also encompass details and analysis of the processing conditions, material microstructure and other sample properties.

Considering the above aspects, although major RFQ design changes will not be needed, in-depth multi-physics analysis of the RFQ system will be soon needed (e.g. thermal expansion, material tensions, conductivity etc.). Subsequently, the cooling connectors and adaptors should be added to the next design iterations, also implementing improvements in cooling channel shape and their internal surface conditioning, optimization of lattice structure size and shape. Empirical experience showed that in the future, it is recommended to use solid supports in the region of the vane tip or throughout the bottom of the RFQ.

Finally, various hybrid machining, post-processing and surface finishing scenarios are being considered (e.g. laser smoothening, abrasive-vibro-processing, high-speed milling [14, 15] etc.). Utilisation of the 3D surface roughness parameters and surface texture standards as per ISO 25178 could be beneficial for the better understanding of the RFQ surface microtopography and would help to quantify the influence of the technological AM parameters to the final surface roughness quality.

# 6      Acknowledgements

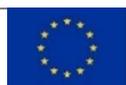 This project has received funding from the European Union's Horizon 2020 Research and Innovation programme under grant agreement No 101004730